\documentclass[12pt]{article}

\catcode`\@=11

\global\arraycolsep=2pt
\usepackage{amsbsy,amssymb,latexsym,amsfonts,amsmath}
\usepackage{graphicx,color}
\usepackage{feynmf}
\allowdisplaybreaks

\begin{document}
\begin{flushright}
\parbox{4.2cm}
{RUP-21-1}
\end{flushright}

\vspace*{0.7cm}

\begin{center}
{ \Large On the trace anomaly of Chaudhuri-Choi-Rabinovici model}
\vspace*{1.5cm}\\
{Yu Nakayama}
\end{center}
\vspace*{1.0cm}
\begin{center}

Department of Physics, Rikkyo University, Toshima, Tokyo 171-8501, Japan

\vspace{3.8cm}
\end{center}

\begin{abstract}
Recently a non-supersymmetric conformal field theory with an exactly marginal deformation in the large $N$ limit was constructed by Chaudhuri-Choi-Rabinovici. On a non-supersymmetric conformal manifold, $c$ coefficient of the trace anomaly in four dimensions would generically change. In this model, we, however, find that it does not change at the first non-trivial order given by three-loop diagrams.
\end{abstract}

\thispagestyle{empty} 

\setcounter{page}{0}

\newpage

In four-dimensional conformal field theories, the trace anomaly has the form
\begin{align}
T^{\mu}_{\mu} = c \mathrm{Weyl}^2 - a \mathrm{Euler}
\end{align}
and it is known that coefficient $a$ cannot change under exactly marginal deformations, but coefficient $c$ may \cite{Osborn:1991gm}\cite{Nakayama:2017oye}\cite{Meltzer:2017rtf}\cite{Bzowski:2018fql}\cite{Solodukhin:2019xwx}\cite{Niarchos:2019onf}\cite{Niarchos:2020nxk}. However, there has been no explicit field theory example where $c$ changes (except for the effective holographic constructions in \cite{Nakayama:2017oye}). The main obstruction has been that we have no good examples of non-supersymmetric conformal field theories with exactly marginal deformations; in superconformal field theories, while it is easier to realize exactly marginal deformations, $c$ does not change \cite{Anselmi:1997am}.

Recently, Chaudhuri-Choi-Rabinovici have constructed a non-supersymmetric conformal field theory with an exactly marginal deformation in the large $N$ limit \cite{Chaudhuri:2020xxb}.\footnote{See also \cite{Chai:2020onq}\cite{Chai:2020zgq} for other recently constructed examples of non-supersymmetric field theories with exactly marginal deformations in different dimensions than four.} This theory may serve as a first non-trivial check if $c$ can really change under exactly marginal deformations. In this short note, we, however, show that it does not change at the first non-trivial order given by three-loop diagrams.

The model (called complex bifundamental model in  \cite{Chaudhuri:2020xxb}   is given by four $SU(N_c)$ gauge theories with names $1,1',2$ and $2'$, each of which has $N_f$  Dirac fermions in the fundamental representation. We have two complex scalars in the bifundamental representations $\Phi_1$ (under gauge group $1$ and $1'$) and $\Phi_2$ (under gauge group $2$ and $2'$). It has no Yukawa interaction, absence of which is protected by chiral symmetry, but it has a scalar potential
\begin{align}
V &= \tilde{h}_1 \mathrm{Tr}[\Phi_1^\dagger \Phi_1 \Phi_1^\dagger \Phi_1] +  \tilde{h}_2 \mathrm{Tr}[\Phi_2^\dagger \Phi_2 \Phi_2^\dagger \Phi_2] \cr
&+  \tilde{f}_1 \mathrm{Tr}[\Phi_1^\dagger \Phi_1]\mathrm{Tr}[\Phi_1^\dagger \Phi_1] + \tilde{f}_2 \mathrm{Tr}[\Phi_2^\dagger \Phi_2]\mathrm{Tr}[\Phi_2^\dagger \Phi_2]  +  2 \tilde{\zeta} \mathrm{Tr}[\Phi_1^\dagger \Phi_1]\mathrm{Tr}[\Phi_2^\dagger \Phi_2] \ . 
\end{align} 
We take the Veneziano limit of $N_c, N_f \to \infty$ with fixed $x=\frac{N_f}{N_c}$ and consider the limit $x \to \frac{21}{4}$ to make the theory weakly coupled. 

In terms of rescaled coupling constants ($i=1,2$)
\begin{align}
\lambda_i = \frac{N_c g_i^2}{16\pi^2} \ , \ h_i = \frac{N_c \tilde{h}_i}{16\pi^2} \ , \ f_i = \frac{N_c^2 \tilde{f}_i}{16\pi^2} \ , \ {\zeta} = \frac{N_c^2 \tilde{\zeta}}{16\pi^2} \ ,
\end{align}
the renormalization group beta functions in the Veneziano limit are expressed  as (no sum over $i$ unless explicitly shown)
\begin{align}
\beta_{\lambda_i} &= -\frac{21-4x}{3} \lambda_i^2 + \frac{-54+26x}{3} \lambda_i^3 \cr
\beta_{h_i} &= 8h_i^2 - 12 \lambda_i h_i + \frac{3}{2}\lambda_i^2 \cr
\beta_{f_i} & = 4 f_i^2 + 16 f_i h_i + 12 h_i^2 + 4\zeta^2 -12 \lambda_i f_i +\frac{9}{2}\lambda_i^2 \cr
\beta_{\zeta} & = \zeta \sum_{i=1}^2 (4f_i + 8h_i - 6\lambda_i) \ .
\end{align}
The zero of the beta functions was studied in \cite{Chaudhuri:2020xxb} and they found that there exists a conformal manifold given by 
\begin{align}
\lambda_1 &= \lambda_2 = \lambda = \frac{21-4x}{-54+26x} \cr
h_1 & = h_2 = \frac{3-\sqrt{6}}{4}\lambda \cr
f_p & \equiv \frac{f_1+f_2}{2} = \sqrt{\frac{3}{2}} \lambda \cr
\zeta^2 + f_m^2 &= \frac{18\sqrt{6}-39}{16} \lambda^2 \ \ , \label{conf}
\end{align}
where $f_m \equiv \frac{f_1 - f_2}{2}$. From the last line of \eqref{conf}, we see that it has the topology of a circle. As long as $\lambda$ is small, we may neglect higher order corrections.

We now ask if the coefficient $c$ in the trace anomaly can change on this conformal manifold. In addition to the coupling constant independent contributions from the one-loop diagrams (that count a number of fields), the coupling constant dependent contributions to the trace anomaly that are relevant for us come from the three-loop diagrams shown in Fig 1. The detailed computation for diagram (A) (as well as  other two-loop diagrams) can be found in \cite{Jack:1983sk}\cite{Jack:1984sq}\cite{Jack:1985wd},\footnote{The three-loop diagrams of (B)(C)(D) are not evaluated in the literature, but we see that diagram (B) and (C) do not contribute to $c$. Diagram (D) may contribute in general, but the contributions to $c$ in our theory do not depend on $\zeta$  or $f_m$  from the symmetry of the diagrams. } but we only need the relative coefficient, so we can simply work on combinatorics.

Up to an overall proportionality factor, the result in the Veneziano limit is summarized as
\begin{align}
c_{\text{2,3-loop}} =- 4 f_m^2  - 4 \zeta^2 + c_\lambda \lambda^2 \ 
\end{align}
on the conformal manifold, where $c_\lambda$ is some numerical constant, which is unimportant for our discussions.\footnote{A typo in the two-loop gauge contribution \cite{Jack:1985wd} that could affect $c_{\lambda}$ has been corrected in \cite{Osborn:2016bev}.} Since the relative coefficient appearing here coincides what appears in the last line of \eqref{conf}, we conclude that $c$ does not change on the conformal manifold although the value itself is perturbatively corrected. We also note that these two- and three-loop diagrams do not change the value of $a$ as anticipated \cite{Osborn:1991gm}\cite{Komargodski:2011vj} (rather trivially without cancellation unlike $c$).

The result is surprising in the sense that we generically expect that $c$ would change on non-supersymmetric conformal manifold. It is an interesting question to see if the higher loop corrections modify our conclusion. It may be possible to relate the all-loop argument for the existence of the exactly marginal deformation in \cite{Chaudhuri:2020xxb} with the computation of $c$ by closing all the external lines in beta functions to make vacuum diagrams. 

\begin{fmffile}{diagram}


(A)
\begin{fmfgraph}(125,125)
\fmfleft{i}
\fmfright{o}
\fmf{phantom,tension=10}{i,v1}
\fmf{phantom,tension=10}{v2,o}
\fmf{dashes,left,tension=0.4}{v1,v2,v1}
\fmf{dashes,left=0.5}{v1,v2}
\fmf{dashes,right=0.5}{v1,v2}
\end{fmfgraph}

(B)
\begin{fmfgraph}(125,125)
\fmfleft{i}
\fmfright{o}
\fmf{phantom,tension=10}{i,i1}
\fmf{phantom,tension=10}{o,o2}
\fmf{dashes,left,tension=0.4}{i1,v1,i1}
\fmf{dashes,right,tension=0.4}{o1,v1,o1}
\fmf{photon,right,tension=0.4}{o1,o2,o1}
\end{fmfgraph}

(C)
\begin{fmfgraph*}(125,125) 
\fmfleft{i}
\fmfright{o}
\fmftop{t1,t2,t3}
\fmfbottom{b1,b2,b3}
\fmf{phantom,tension=5}{i,v1}
\fmf{phantom,tension=5}{o,v2}
\fmf{dashes,right=0.18,tension=3}{v2,vt3}
\fmf{dashes,right=0.18,tension=3}{vt3,vt2}
\fmf{dashes,right=0.18,tension=3}{vt2,vt1}
\fmf{dashes,right=0.18,tension=3}{vt1,v1}
\fmf{dashes,right=0.18,tension=3}{v1,vb1,vb2,vb3,v2}
\fmf{photon,tension=0.5}{v2,v1}
\fmf{dashes,right,tension=0.05}{vb2,b4,vb2}
\fmf{phantom}{t1,vt1}
\fmf{phantom}{t2,vt2}
\fmf{phantom}{t3,vt3}
\fmf{phantom}{b3,vb3}
\fmf{phantom}{b2,vb2}
\fmf{phantom}{b1,vb1}
\fmf{phantom,tension=5}{b4,b2}
\end{fmfgraph*} 

(D)
\begin{fmfgraph}(125,125)
\fmfleft{i}
\fmfright{o}
\fmftop{t}
\fmf{phantom,tension=10}{i,i1}
\fmf{phantom,tension=10}{o,o1}
\fmf{phantom,tension=3}{t,t1}
\fmf{dashes,left=0.5,tension=1}{i1,v1,v2,v3,o1}
\fmf{dashes,left=0.5,tension=1}{o1,v2,i1}
\fmf{photon,left=0.3,tension=1.5}{v1,t1,v3}
\end{fmfgraph}

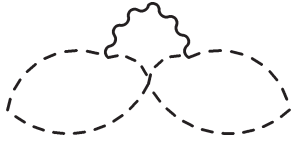
Fig 1: Three-loop Feynman diagrams that could contribute to $c$.

\end{fmffile}

\section*{Acknowledgements}
This work is in part supported by JSPS KAKENHI Grant Number 17K14301. It is motivated from the online talk by Z.~Komargodski at YITP workshop on Strings and Fields 2020, which the author watched on Youtube later.

\end{document}